\documentstyle[preprint,aps,prb]{revtex}

\newcommand{\mb}[1]{ {\mbox{\boldmath{$#1$}}}  }

\begin{document}
\draft

\title{Percolation of Superconductivity.}

\author{G. Litak}
\address{Department of Mechanics,
 Technical University of Lublin\\
Nadbystrzycka 36, PL--20-618  Lublin, Poland}

\author{B.L. Gy\"{o}rffy}
\address{H.H. Wills Physics Laboratory, University of
Bristol \\ Tyndall Ave, Bristol BS8 1TL, United Kingdom
}

\date{\today}
\maketitle

\begin{abstract}

\noindent 
In case of superconductors whose electrons attract each other only if
they are near certain centers, the  question  arises  'How many
such centers are needed to make the ground state superconducting?'
We shall examine it in the context of a random $U$
Hubbard model. In short we study the case where $U_i$ is $-|U|$
and 0 with probability $c$ and $1-c$ respectively on a lattice 
whose sites are labelled $i$
using the Gorkov decoupling and the Coherent Potential Approximation 
(CPA). We argue that for this model there is a critical concentration
$c_0$ below which the system is not a superconductor.  
\end{abstract}

\section{introduction}  

In many attempts to construct a viable model for High 
Temperature Superconductors the notion of negative -- $U$ centers
is invoked \cite{Don90,Wil89,Mic90,Lev94}. In this connection there is a
simple, natural
question that arises: How many such centers are needed to make
a superconductor. In this contribution we shall argue that under certain
circumstances there is a
critical concentration $c_0$ below which there is no superconducting
order. Moreover, we developed a strategy for investigating the factors
which determine $c_0$.

In order to deal with a well posed problem we shall study a Random -- $U$
Hubbard Model defined by the Hamiltonian

%1
\begin{equation}
H= -\sum_{i j \sigma}t_{i j} c_{i \sigma}^{+} c_{j \sigma} +
\sum_{i \sigma}  U_i  c_{i \sigma}^{+} c_{i \sigma} c_{i -\sigma}^{+}
c_{i -\sigma} - \mu\sum_{i \sigma} c_{i\sigma}^+c_{i\sigma}~ , \label{E1}
\nonumber
\end{equation}

where the coupling constant

\begin{displaymath}
U_{i}= \left\{ \begin{array}{rl} -|U| & {\rm with~ probality}~ c 
\\ 0  & {\rm with~ probability}~ 1-c   
\end{array}~. \right.
\end{displaymath}

The question we shall ask is: Is there a finite concentration $c_0$
such that for $c < c_0$ the cofigurationally averaged, superconducting
long range order parameter $\overline{\chi}$ vanishes even at zero
temperature?

As is natural we define $\overline{\chi}$ by the relation

%2
\begin{equation}
\overline{\chi}= \frac{1}{N} \sum_i \chi_i~, \label{E2}
\end{equation}

\noindent where the local pairing amplitude is given by

%3
\begin{equation}
 \chi_{i}=< c_{i \uparrow} c_{i \downarrow}>~. \label{E3}
\end{equation}

Following the conventional notation the bracket $<...>$ denotes a
thermodynamic average and
$\overline{\Theta}$, for an arbitrary operator $\hat{\Theta}$,  implies
the average of $\hat{\Theta}$ over all configurations ${U_i}$, 
such that the fraction
of negative $U$ sites is c, with equal weight. 
A sample will be said to be superconducting if $\overline{\chi} \neq 0$
This implies that $\chi_i \neq 0$ on a finite fraction of all sites.
Namely, if $\chi_i \neq 0$ only on a finite number of sites
$\overline{\chi}$ will go to zero as $N \rightarrow \infty$ and the system
will be regarded as not superconducting.

To make progress we
calculate $\chi_i$ within the Hartree -- Fock -- Gorkov (HFG)
Decoupling  scheme
for the Greens functions and the averaging over the $U$--configurations
is accomplished with the help of the
Coherent Potential  Approximation (CPA) \cite{Gon92}. In short, at the
risk of
missing some important feature of the problem, like localization of
electrons, we develop a mean field theory for the phenomena described by
$H$ in Eq. (\ref{E1}) This approach may be
justified by noting that little is known about the problem at hand
systematically \cite{Don90,Wil89,Mic90} and hence as a preliminary study a
mean
field theory
is
called for.

Note that the simplest approximations to the problem would be to set 
$U_i$ at each site equal to its average  value  $\overline{U}=cU$.  In 
some contexts 
this is called the Virtual Crystal Approximation \cite{Gyo79,Gon92}.
Since, as
is well
known \cite{Mic90}, any amount of 
attraction leads to superconductivity, $\overline{U}=cU$ implies
superconductivity 
for all
non zero concentrations  with the transition temperature $T_C$ decreasing
albeit non-analytically,  with
$c$.
Thus  before setting out the details of the above theory it is worthwhile to 
pause,
briefly, to consider a number of fairly general  arguments  which 
suggest
that the above conclusion is premature and that there is a critical
concentration $c_0$ of negative $U$ -- centers for
superconductivity. \\ \\

\noindent
{\bf i. Classical Percolation Theory} for a mixture of two metals with
resistivities $\rho_1$ and $\rho_2$ have been studied in the Effective 
Medium Approximation \cite{Bru35,Gyo79,Kir79}. For $\rho_1=\rho_0$ and
$\rho_2=0$ (Fig. 1), namely
in the case where metal 2 is a superconductor, it yields
an effective resistivity given by

%4
\begin{equation}
\rho_{eff}= \left\{  \begin{array}{rl} \rho_0 (1-dc) & {\rm for}~ c <
c_0=\displaystyle{\frac{1}{d} }
\\ ~~ \\ 0  & {\rm for}~ c > c_0= \displaystyle{\frac{1}{d}}  
\end{array} \right. ~, \label{E4}
\end{equation}

where $d$ is the number of spatial dimension in which percolation is
allowed. Thus, this model predicts critical concentrations in $d=1,2$ and
3 dimensions,
More  over,  $c_0$  depends  on  the  dimensionality  $d$.   More 
generally
$\rho_{eff} \sim (c-c_0)^s$ near $c_0$ but a mean field theory cannot
 be expected
to deal with the critical exponent $s$ adequately. \\ \\

\noindent
{\bf ii. The propagation of Cooper pairs} between negative $-U$ centers
by hopping from site to site, where $U=0$ on the intermediate sites, is
depicted in Fig. 2.
Assuming that the distance between two negative--$U$ centers is 
$c^{-1/d}$,
in units of the lattice constant $a$ on a $d$ dimensional lattice, we
estimate the number of individual hops $l$ necessary to reach one such
center from its nearest neighbors. Assuming random walk
$c^{-1/d}=l^{1/(d-1)}$ (for $d > 1$). If each hop takes $\hbar/W$ 
seconds where $W$ is the bandwidth
for the Cooper pairs, the time to travel between two negative $U$ centers
is given by $\tau=(\hbar/W)~c^{-(d-1)/d}$. 
Now we note that in between two centers the Cooper
pair is without its binding energy $U$. Consequently, such travel is
allowed only for such times $\delta t$ that the energy uncertainty 
$\delta E =\hbar/\delta t > U$. Taking $\delta t=\tau$ we conclude that for 
$\delta E =W c^{(d-1)/d} > U$ the pair will propagate for
$\delta E =W c^{(d-1)/d} < U$ the pair will not propagate. Thus for
$c < c_0$, where

%5
\begin{equation}  
c_0=\left( \frac{U}{W} \right)^{\displaystyle\frac{d}{d-1}}~,
\label{E5}
\end{equation}
 
\noindent a system of negative $U$ centers will not be superconducting.
Presumably, the Cooper pairs will be localized. On the other hand for 
$c > c_0$ it will be a superconductor. \\ \\

\noindent
{\bf iii. Localization of Cooper pairs by local charge and
order -- parameter -- phase fluctuation} is the third argument which we
wish to recall briefly. It was explored in the present context by
Doniach and Inui \cite{Don90}. In a Ginzburg Landau theory on a lattice
the
relative phases of the local order parameters $\psi_i=|\psi_i|\rm{e}^{i
\Theta_i}$
are determined by the quadratic term in the free energy function 
$F(\{ \psi_i \})$. 
This  may be written in the form of Josephson coupling energies
$F( \{ \Theta_i \})=\frac{1}{2} \sum_{ij} E^J_{ij} \cos (\Theta_i-\Theta_j)$,
where the precise relationship of the coefficients  to
various parameters of the theory need not concern us here.
To describe charge fluctuations associated with Cooper pairs arriving
and leaving
a site a charging energy needs to be added to $F$. Because, the local
potential is related to the phase by Josepson voltage relation $V_i=
\frac{\hbar}{2 \rm{e}} \dot{\Theta}_i$ 
this has the form of a kinetic energy term. Finally, to recover a
microscopic description the local phases are treated as quantum
mechanical variables by the Hamiltonian

%6
\begin{equation}
H= \frac{1}{2} E_C \sum_i \left( i \frac{\partial}{\partial \Theta_i}
\right)^2 -\frac{1}{2} E_J \sum_{ij} \cos(\Theta_i -\Theta_j)~.
\label{E6}  
\end{equation}

This is a much studied Hamiltonian in connection with granular
superconductivity. In particular it was investigated by Gosset and
Gy\"{o}rffy \cite{Gos91} in the Hartree -- approximation. In short they
factorized the
wave -- function as shown below

%7
\begin{equation}
\Psi(\{\Theta_i\})=\prod_i\phi_i(\Theta_i)
\label{E7}
\end{equation}
and found the following self--consistent equation for the individual
site
wave function $\phi_0(\Theta)$

%8
\begin{equation}
\left[ - \frac{1}{2} E_C \left( \frac{\partial}{\partial
\Theta}\right)^2 - E_J \cos(\Theta-\overline{\Theta}) \right]
\phi_0(\Theta)=
E_0\phi_0(\Theta)~, \label{E8}
\end{equation}
where
$$
\left< e^{i\Theta} \right> = \int \rm{d} \Theta \phi_0^*e^{\rm{i}
\Theta} \phi_0(\Theta) = \rho e^{\rm{i} \overline{\Theta}} 
$$

The amplitude $\rho$, determined by solving the above equation
numerically is shown in Fig. 1 (in Ref. \onlinecite{Gos91}),as a function
of the ratio $E_J/E_C$
($\equiv$ Josephson energy/charging energy) For $E_J/E_C < 0.125$ we find
$\rho=0$ and hence we conclude that the system of point superconductors
we have been considering do not have long range superconducting order.
Clearly, it is tempting to  associate  $E_J$  with  the  coupling 
between the
negative -- $U$ centers in our Hubbard model and assume that it goes to
zero
as $c \rightarrow 0$. Evidently, this would imply a critical concentration
determined by $E_C=E_J(c_0)$ . In short, charge fluctuations can destroy the
phase coherence of superconducting order parameter if the coupling
between the negative -- $U$ centers drops below certain critical value. 
Indeed this was one of the main point of the paper by Doniach and Inui
\cite{Don90}. In what follows we shall develop a strategy for
investigating the
link between the microscopic model defined by Eq. (\ref{E1}) and the above
semi-phenomenological arguments. 

In concluding this introduction we note that the specific task we shall
undertake is a contribution to the general problem of treating disorder
and electron--electron interactions simultaneously For a comprehensive
discussion of the relevant issues in this field the reader is referred to
the relatively recent review article by Belitz and Kirkpatrick
\cite{Bel94}.

\section{The Coherent Potential Approximation for the Random--$U$
Hubbard Model} 

The physics described by this simple model appears to be exceedingly
rich. For instance, one might expect that, under some circumstances,
the Cooper pairs are subject to Anderson localization \cite{Tho78} and
hence
they
form a random set of Andreev scatterers for the quasi--particles
\cite{Lam91}.
Such
system of scattering centers may then Anderson localize the
quasi--particles themselves and turn the system into an insulator below
the critical concentration $c_0$ for superconductivity. 
However very little systematic fully microscopic work has been done on the
problem and
hence, as a preliminary exercise, a mean field theoretic treatment is
called for even at the risk of failing to capture some of its
important features. In any case, as we shall show, even such limited
description turns out to be of physical interest.

Formally, the task is to find the Greens function

%9
\begin{equation}
\mb G(i,j;\tau;\{U_i\})= - \left[\begin{array}{cc}
< T \{ c_{i\uparrow}(\tau) c^{+}_{i \uparrow}(0) \} > & 
<T \{c_{i \uparrow}(\tau) c_{i \downarrow}(0)\}> \\    
< T \{ c^{+}_{i\downarrow}(\tau) c^{+}_{i \uparrow}(0) \} > &  
<T \{c_{i \downarrow}(\tau) c_{i \downarrow}(0)\}>  \end{array} 
\right]~,   \label{E9}
\end{equation}

\medskip

\noindent where the creation and annihilation operators
$c_{i \sigma} (\tau)$ and $c_{i \sigma}^+(\tau)$ evolve 
in complex time $\tau$ according the random--$U$
Hamiltonian $H$ in Eq. (\ref{E1}), $T$ is the $\tau$--ordering operator,
brackets
$<...>$  denote here
the usual equilibrium thermal averages corresponding to $H$,
and average the result with respect to
all arrangement of the $U$-centers each denoted by $\{U_i\}$.
In short we wish to find

%10
\begin{equation}
\overline{\mb G}(i,j;\tau)= \sum_{\{U_i\}} P(\{U_i\})
\mb G(i,j;\tau;\{U_i\})~~, \label{E10}
\end{equation}

\noindent where the probability distribution is assumed to be of form

%11
\begin{eqnarray}
P(\{U_i\})&=& \prod_i P(U_i)  \label{11} \\
{\rm with~~} P(U_i) &=& \left\{ \begin{array}{r} c~~ {\rm for}~~ U_i = U
\\
1-c~~ {\rm for}~~ U_i=0 \end{array} \right.~. \label{E12}
 \end{eqnarray}

Note that the local order parameter defined by Eq. (\ref{E3}) is given by

%12
\begin{equation}
\overline{\chi_i}=\overline{G}_{12}(i,i; \tau=0^+)~~, \label{E13}
\end{equation}

\noindent and hence the knowledge of the averaged one particle Greens functions
matrix
is sufficient to address the question whether or not there is
superconducting long range order at a given concentration $c$.

As we have indicated above we shall now proceed to a mean field
approximation to the above problem. This consists of two steps. Firstly,
we
make use of the Hartree--Fock--Gorkov decoupling scheme to find the
following 'mean-field' equation of motion

%13
\begin{equation}
\sum_{l} \left[ \begin{array}{c} (\omega_n+\mu-\frac{1}{2}U_{i}n_{i})
\delta_{il} +t_{il}~~~~~\Delta_{i} \delta_{il} \\
\Delta^{*}_{i}\delta_{il}~~~~~(\omega_{n}-\mu+\frac{1}{2}U_{i}n_{i})
\delta_{il}-t_{il} \end{array}
\right] \mb G(l,j;\omega_{n})=
\mb 1\delta_{ij}~,
\label{E14}
\end{equation}
where
%14
\begin{eqnarray}
n_{i} & =&  \frac{2}{\beta} \sum_{n} {\rm e}^{{\rm i} \omega_{n} \delta}
G_{11}(i,i;\omega_{n})~, \nonumber \\
\chi_{i} & =&  \frac{1}{\beta} \sum_{n} {\rm e}^{{\rm i} \omega_{n} \delta}
G_{12}(i,i;\omega_{n})~, \nonumber \\
\Delta_{i} & = & -U_{i} \chi_{i} ~.  \nonumber
\end{eqnarray}

Secondly, we find average of the solution to Eq. (\ref{E13}), namely 
$\mb G(i,j;\omega_n;\{u_i\})$,
over all $U$--configurations using the Coherent Potential
Approximation
(CPA). The justification for this second step is that the CPA is well
known to be a reliable mean--field theory of disorder for wave
propagation in a medium
described by independent random variables \cite{Bru35,Gyo79,Kir79,Vla92}.

To implement the CPA we rewrite Eq. (\ref{E13}) in the Dyson form

%15
\begin{equation}
\mb G(i,j;\omega_n)=\mb G^0
(i,j;\omega_n)+ \sum_{l} \mb G^0(i,l;\omega_n)~
\mb V_l~ \mb G(l,j;\omega_n)~,
\label{E15}
\end{equation}
where
%16
\begin{equation}
V_l = \left( \begin{array}{cc} \frac{1}{2} U_{l}n_{l} & -\Delta_l \\
     -\Delta_l^{*} & - \frac{1}{2} U_{l} n_{l} \end{array} \right)~.
\label{E16}
\end{equation}

The CPA recipe for $\overline{\mb G}(i,j;\omega_n)$ 
 is to set it equal to the coherent Greens
function $\mb G^C(i,j;\omega_n)$ which is the
 solution of Eq. (\ref{E15}) for the case where the
random potential $\mb V_l$ 
is replaced by the energy dependent, complex
coherent potential $\mb \Sigma(\omega_n)$,
the same on every site. To determine the coherent potential (self--
energy) we study, in turn a $U_i=-|U|$ impurity in the coherent lattice.
On the impurity
site at $i$ we find

%17
\begin{equation}
\mb G^{\alpha}(i,i;\omega_n)= \left[ 1 - 
\mb G^C(i,i;\omega_n)
\mb V_i^{\alpha}  - \Sigma(\omega_n) \right]^{-1}
\mb G^C(i,i;\omega_n)~,~~~ {\rm for}~~\alpha=0~~ 
{\rm
and}~~ U~,
\label{E17}
\end{equation}

\noindent where

%18
\begin{equation}
\mb V^{\alpha=0}_i ~~~~{\rm and}
~~~~\mb V^{\alpha=U}_i = \left( \begin{array}{c}
\frac{1}{2} Un_i~~~~-\Delta_i \\
-\Delta_i^*~~~~-\frac{1}{2}Un_i \end{array} \right)~.
\label{E18}
\end{equation}

\noindent Then, the usual CPA condition which determines the self--energy 
$\mb \Sigma(\omega_n)$
is given by 

%19
\begin{equation}
c \mb G^{(0)} (i,i;\omega_n) + (1-c)
\mb G^{(U)}(i,i;\omega_n) =
\mb G^C(i,i;\omega_n)~.
\label{E19} 
\end{equation}

%A pictorial illustration of this basic equation of the theory is given
%in Fig. 2. 
Similar equations have been used to describe random
superconductors by Lustfeld \cite{Lus73} and more recently by Litak
{\em et al} \cite{Lit92}.
The principle difference between our present concerns and that of
these earlier authors is that we are focusing on the randomness of the
interaction parameter $U_i$ and not on the random site energies
$\epsilon_i$
as was their aim. To put it on other way we are studying a problem
analogues to that of 'spin--glass' \cite{Lev94} rather than that of dirty
superconductors.

Equations (\ref{E17},\ref{E18}) and Eq. (\ref{E19}) together with

%20
\begin{eqnarray}
n^{\alpha} & = & \frac{2}{\beta} \sum_n \rm{e}^{i \omega_n \delta}
G_{11}^{ \alpha} \nonumber \\
\chi^{\alpha} & = & \frac{2}{\beta} \sum_n \rm{e}^{i \omega_n \delta}
G_{12}^{ \alpha} \label{E20} \\
\Delta^U & = & -U^{\alpha} \chi^{\alpha} \nonumber \\
\overline{n} & = & cn^{(U)} + (1-c) n^{(0)}  \nonumber \\
\overline{\chi} & = & c \chi^{(U)} + (1-c)\chi^{(0)}~, \nonumber
\end{eqnarray}

\noindent where $\alpha=0$ and $U$ as before (Eq. \ref{E17}), are the
fundamental 
equations of our
theory. 

\noindent Manipulating the $CPA$ equations yield the following gap
equation:
\bigskip

%21
{\normalsize
\begin{eqnarray}
\overline\chi &=&- \frac{U}{\beta} \sum_{n} {\rm e}^{ {\rm i} \omega_{n}
\delta}
\left[ - \frac{c}{2 \omega_{n}} {\rm Tr} \left\{ \mb G^{(U)}
\right\} \frac{ \| \mb G^{(U)} \| }{
\| \mb G^{C} \| } +\| \mb G^{(U)} \|
\left( c \frac{ 2 \omega_{n} - {\rm Tr} \mb \Sigma 
}{2 \omega_{n}}   \frac{\|\mb G^{(U)}\|}
{ \| \mb G^{C}\|} -1 \right) \right] \overline \chi~.
\nonumber \\
 &~& \label{E21}
\end{eqnarray}
}
\bigskip

In what follows we present results of solving the above equations numerically 
for various
interesting regimes.
Of particular interest is the large $U$ limit. As $U_i$ change
its values form 0 to $-|U|$ there exists Mott--Hubbard metal--insulator transition
 for large enough interaction 
$|U|$. An other interesting feature of the problem at hand
is that  fluctuations of
pairing potentials $\Delta_i$, which changes randomly from 0 to
$\Delta^U$,
invalidate the Anderson theorem \cite{And59,Gyo97,Lit98} and hence states
appear in the gap \cite{Lit99}.  
 
\bigskip

\noindent

\section{ORDER PARAMETER FLUCTUATIONS}

At first we have   calculated $T_C$ and $\overline{\chi}$ for zero
temperature 
$(T=0)$ by means of
 VCA where, as we mentioned in the introduction effective interaction
between electrons 
 $U_{eff}=cU$. 
Figs. 3$a$ and 3$b$ show the critical 
temperature $T_C^{(c)}$  normalized to the
corresponding
quantities 
of the clean system with $U$ on every site, namely $T_C(c=1)$ and averaged
pairing
parameter 
 $\overline{\Delta}(T=0,c=1)$, calculated for effective interaction
$U_{eff}$
respectively as functions of concentration $c$. Calculations were done
for various values of the interaction parameter $U/W$:
-0.3, -0.5, -1.0, -2.0
and
 for a half filled band: $n=1$.
One can see that for these approximations there is no evidence of 
critical concentration $c_0> 0$ below which the systems is normal at $T=0$ 
i.e. no percolation. 

As is clear from Eq. (\ref{E14}), $U_i$ fluctuating between 0 and $-|U|$
has two distinct direct consequences. On the one hand it causes the
Hartree 
potential $\frac{1}{2} U_in_i$ to fluctuate. On the other it gives rise to
a fluctuating 
pairing parameter -- $\Delta_i$. As it turns out these two effects have very different 
influence on the solutions to Eqs.
(\ref{E18},\ref{E19},\ref{E20},\ref{E21}). Therefore,
we 
examine them
separately.
As disorder was treated by CPA,
at first we made calculations after neglecting Hartee potential
$\frac{1}{2}U_in_i$, in Eq. (\ref{E18})
 and studied the case of order parameter fluctuation on their own.
This means that we took the impurity potential in Eq. (\ref{E18}) to be 

%21
\begin{equation} \mb V_l = \left( \begin{array}{cc} 0 &
-\Delta_l \\
     -\Delta_l^{*} & 0 \end{array} \right)~.
\end{equation}

In Figures 4$a$, $b$, $c$ we show 
the critical temperature $T_C/T_C(c=1)$ ($a$), the order
parameter (for $T=0$) $\overline \chi/\chi(c=1)$ ($b$)  and the local
pairing potential
on $U$ site $\Delta_U/\Delta_U(c=1)$ ($c$), versus concentration 
of negative centers $c$ for $n=1$.
Calculations were done 
by means of CPA neglecting the Hartee term and using
 the same values of 
interaction parameters as in Fig. 3.

Surprisingly, our simplified $CPA$ results  agree with the  $VCA$
argument  in 
the introduction in as much as we found non--zero local order  parameter 
on both the 
$U=0$
and $U<0$ sites at all concentrations $c \neq 0$. That is to say
 we obtained finite $\overline{\chi} \neq 0$ and $T_C$ for any value of 
concentration 
$c$ and interaction $U_i < 0$ and no evidence of percolation.
The order parameter $\overline{\chi}$ increases gradually from 0 to 
its maximal value with
changing the concentration
$c$. 
Interestingly,   in the large $U$ limit, $|U|/W > 0.5$,  $T_C$ and
$\Delta_U$ 
are nearly constant
for various $c$ and they reach large finite values for arbitrary small concentrations
of negative centers  $c$.

To return to the  full CPA solution we have used the full impurity
potential 
$\mb V_i^{\alpha=U}$ as in Eq. (\ref{E18}). Figures 5$a$, $b$, $c$, $d$
show
the critical temperature $T_C^{(c)}/T_C(c=1)$ ($a$), the order
parameter $\overline \chi/\chi(c=1)$ ($b$) and the local pairing potential
on $U$ site $\Delta_U/\Delta_U(c=1)$ ($c$), versus concentration 
of negative centers $c$ for $n=1$ and the same interactions as in Figs 3 and 4.
Here one can clearly see that  all that quantities $T_C$, $\overline 
\chi(T=0)$
and $\Delta_U$    
are tending to zero for some small enough  concentration of negative $U$ centers
$c_0$. Below this critical concentration the system is normal. 
For larger interaction 
($U/W=$ -1.0, -2.0)
$c_0=0.5$  and  the order parameter 
scales as $\chi \approx (c-c_0)^{1/2}$.
Decreasing $|U|$ we observe systematic decrease of $c_0$. 

To investigate the case of critical concentration $c_0$ we have studied
the density of quasiparticle states both in superconducting and in the
normal states. In the latter case, for large enough interaction $| U| >
0.5 W$, there exists a
band splitting
in the system. With changing the 
concentration $c$ we observe 
Mott metal--insulator transition. It is caused by large fluctuations of 
Hartree term $\frac{1}{2} U_in_i$ (Eq. \ref{E14}) as in the original paper
of
Hubbard \cite{Hub64}.
In Figs. 6 $a$, $b$, $c$ we plotted
the densities of states (full line) and the local density of states on $U$ site
(dashed line) for $U/W=-2.0$, $n=1$ and $T=0$  for different concentrations
$c$: $c=0.4$ ($a$ - normal metal), $c=0.5$ ($b$ - insulator) 
and $c=0.6$ ($c$ - superconductor). The Fermi energy in these plots:
$\epsilon_F=\mu=0$. Thus changing $c$ from 0 to 1 system changes from
normal metal
(Fig. 6$a$)
to a  superconductor  (Fig. 6$c$) through an insulator (Fig. 6$b$).
Remarkably, for a low concentration of negative centers (Fig. 6$a$) 
$c=0.4 < c_0$ (here $c_0=0.5$) in spite of finite and relatively large value
of averaged density of states at the Fermi energy: $\overline D(0)=
-\frac{1}{\pi}
{\rm Im} G_{11}^C(0+{\rm i}\delta)$, the local density of states on $U$
sites
$D^U(0)= -\frac{1}{\pi}
{\rm Im} G_{11}^U(0+{\rm i}\delta)$ appears to be extremely small. 
Evidently the doubly occupied states form a lower 'Hubbard' band split off 
from the upper band which is associated with the singly occupied sites.

This effect has been further investigated for other band fillings.   
The transition from normal to superconducting phase occurred 
for each of band fillings $n$ at some, specific, critical
concentration
$c_0(n)$.
Figures 7$a$, $b$, $c$, $d$ show simultaneously 
the order parameter $\overline\chi$ ($a$), the local pairing potential
on $U$ site $\Delta_U$ ($b$), 
the local charge on $U$ site $n_U$
($c$) and the chemical potential $\mu$ ($d$)
 plotted versus concentration $c$ for $U/W=-2.0$, at $T=0$ 
 and several values of $n$
($n=$ $1.8$, $1.6$, $1.4$, $1.2$, $1.0$, $0.8$, $0.6$, $0.4$, $0.2$)
Interestingly, that transition from superconducting to normal phase
is accompanied by a large value of local charge occupation $n_U$ (Fig. 7$c$)
and large jump
of a chemical potential $\mu$ (from one subband to another)  near $c_0$
(Fig. 7$d$). 
It appears that for $c$ below $c_0$ $n_U \approx 2$. Namely,
every $U$ site is doubly occupied with 
a pair of electrons (Fig. 7$c$). 
Because there are no empty spare $U$ sites in the system these
pairs cannot move. That is to say, they are localized on the $U$ sites.

Similar calculations have been performed for smaller  
interaction $|U|$ ($U/W=-0.5$)
The corresponding results are presented in Figs. 8 $a$--$d$ respectively. 
Here the interaction  $|U|$  
is not large enough to create a band splitting effect but the tendency with
$\overline \chi \rightarrow 0$ is still observable as concentration $c$
 is tending
to some finite $c_0 > 0$ $(c \rightarrow c_0)$.
 Here $c_0$ 
is less that in former case of larger
interaction $|U|=2W$ (Fig. 7). The occupation of negative centers is larger than 
$n$ but clearly less than 2 electrons per site.
For small enough band filling $n=0.2$, the order parameter $\overline 
\chi$ was finite for all
$c$ and we have not observed a percolation phenomenon. 
For larger band fillings we have obtained  $\overline \chi = 0$  below $c <
c_0$
 but instead of a square root behaviour $\overline \chi \approx
(c-c_0)^{1/2}$ for $c$ 
 close to $c_0$ for larger $U$ (U/W=-2.0 fig. 7$a$) here 
 $\chi$ goes to  zero rather in the asymptotic 
way (Fig 8$a$). 

To investigate the demise of the superconducting state near $c_0$ we have
studied the density of states in the appropriate region of parameter
space.
Figures  9$a$ and $b$ shows the quasiparticle densities of states ($a$) 
and the local densities of states on $U$ site ($b$)
for $U/W=-0.5$, $n=0.4$  and  three
values of $c$ specified 
in the figures. It is clearly visible how the superconducting 
gap is filled, due to pair breaking, in with $c$. Beginning from 
the clean system with interaction $U$ in every site $c=1$ we start from 
the sharp edges in the quasiparticle density of states (Fig. 9$a$)
then
for smaller values of $c$ ($c=0.6$) 
the gap parameter $\Delta$ is of the same order (Fig 7$b$) but the real 
gap in the quasiparticle
density of states $\overline D(E)$ changes significantly. The gap becomes
smaller with smaller $c$ and
looses its clear edges.  For small
enough $c$ ($c=0.14$) it nearly disappears. Clearly, the Anderson theorem
for a 
superconuctor  with nonmagnetic disorder is not 
satisfied in this case \cite{And59}. 
As is well known, according to Anderson theorem the gap remains absolute
in presence of disorder due to potential scattering provided the
spatial fluctuations of $\Delta_i$  about $\overline \Delta$ are
negligible \cite{Gyo97}. Clearly, in the random interaction case this is
not true and this kind of disorder leads to pair breaking.

Thus on account for the large fluctuations of 
pairing potential $\Delta_i$ in our system,due to disorder, 
we observe  a qualitative 
change in quasiparticle density of states shown in
Fig 9$a$. These fluctuations lead also to complicated gap equations 
where $T_C$ is determined not only by $\mb G^C$ 
but also by
$\mb  \Sigma$, $\mb G^U$
(Eq. \ref{E21}).

Finally, we investigated the factors which determine the critical
concentration $c_0$.
In Fig. 10 we show $c_0$ as a function of band filling for two interaction
parameters $U/W=-2.0$ and $-0.5$. In both cases function can 
be approximated by  a straight line $c_0=a +b n$. 
In case of $U/W=-2.0$, $a=0$ and $b=0.5$ but for $U/W=-0.5$, $a \approx 0.32$
and $b \approx 0.6 $.

\section{Conclusions} 

\medskip

We have examined the question of percolating superconductivity in the
context of a random $U$
Hubbard model. We have studied the case where $U_i$ is $-|U|$
and 0 with probability $c$ and $1-c$ respectively on a lattice 
whose sites are labelled $i$
using the Gorkov decoupling. Changing concentration $c$ we checked that
simple averaging procedures like Virtual 
Crystal Approximations (VCA) do not lead to any zero temperature 
phase transition. Furthermore, we found that if charge fluctuations are
neglected even a full mean field theory of disorder, like the CPA, does not
predict a percolation transition. However, when the fluctuations in Hartree
potential are included on equal footing with the fluctuations in the
pairing potential $\Delta$ and the problem is treated in the  Coherent
Potential Approximation a percolation phenomena, with a critical
concentration $c_0$ of the negative $U$ centres, is discovered in our
fully
microscopic theory. For $c < c_0$  the lack of superconductivity is due to
Mott localization of Cooper pairs and its high lights the qualitative
difference between disorder in the crystal potential and the disorder in
the interaction between the carriers. Having found the critical
concentration $c_0$ we investigated its dependence on various parameters
which defined the problem. In short we studied $c_0(n,U)$. For strong
attractive interaction $c_0=n/2$ and
$\overline \chi \approx (c-c_0)^{1/2}$ near $c_0$ but for smaller
interaction 
$\overline \chi \rightarrow 0$ (as $c \rightarrow c_0$) rather in
a non polynomial manner. Calculations have been performed by a real space
recursion algorithm which we developed for disordered 
superconductors in earlier publication\cite{Lit95}.

\noindent {\large {\bf Acknowledgements}}

\medskip
  
This work has been partially supported by the
Royal Society. Authors would like to thank prof. 
S. Alexandrov and prof. R. Micnas for helpful discussions.

%\noindent
%{\bf References}

\newpage

%\begin{enumerate} 

\newpage
%\noindent {\bf Figure Captions}

%\begin{itemize}

\begin{figure}
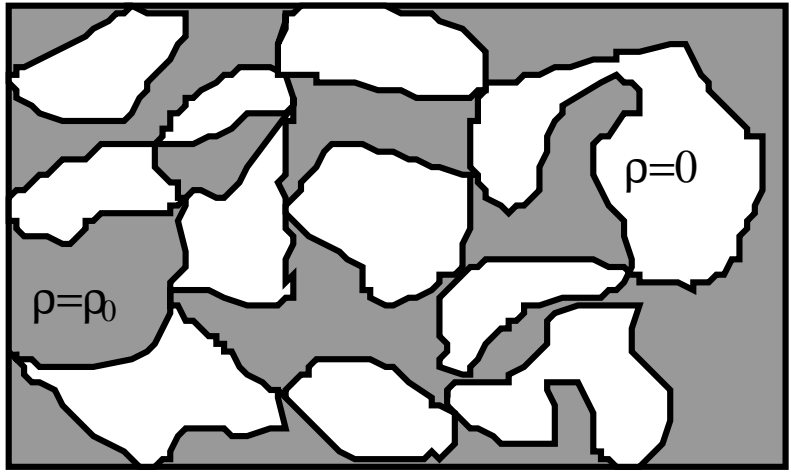

\caption{ A mixture of normal, with resistance $\rho=\rho_0$, and
super,
with $\rho=0$, conductors.
}
\end{figure}

\begin{figure}
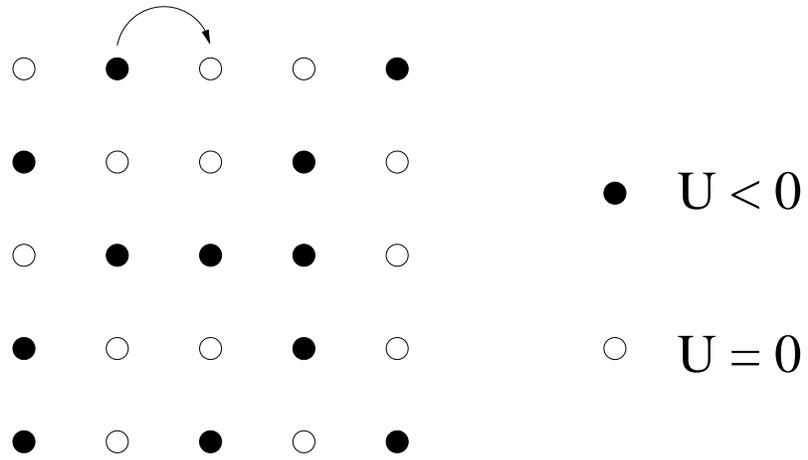

\caption{ The propagation of Cooper pairs between negative $U<0$ centers
by hopping from site to site, where $U=0$ on the intermediate sites.}
\end{figure}

\begin{figure}
\caption{ The critical temperature $T_C/T_C(c=1)$ ($a$) and the pairing
parameter $\bar \Delta/\Delta(c=1)$ ($b$) (calculated with VCA)
versus concentration
of negative centers $c$ for $n=1$. Values of the interaction parameter
$U$ specified
in the figures.}
\end{figure}

\begin{figure}
\caption{The critical temperature $T_C/T_C(c=1)$ ($a$), the order
parameter $\bar \chi/\chi(c=1)$ ($b$)  and the local pairing potential
on $U$ site $\Delta_U/\Delta_U(c=1)$ ($c$), calculated with CPA
neglecting
diagonal Hartree terms, versus concentration
of negative centers $c$ for $n=1$. Values of the interaction parameter
$U$ specified
in the figures.}
\end{figure}

\begin{figure}
\caption{The critical temperature $T_C/T_C(c=1)$ ($a$), the order
parameter $\bar \chi/\chi(c=1)$ ($b$) and the local pairing potential
on $U$ site $\Delta_U/\Delta_U(c=1)$ ($c$),
calculated with CPA including
diagonal Hartree terms, versus concentration
of negative centers $c$ for $n=1$. Values of the interaction parameters
$U$ specified
in the figures.
}  
\end{figure}

\begin{figure}
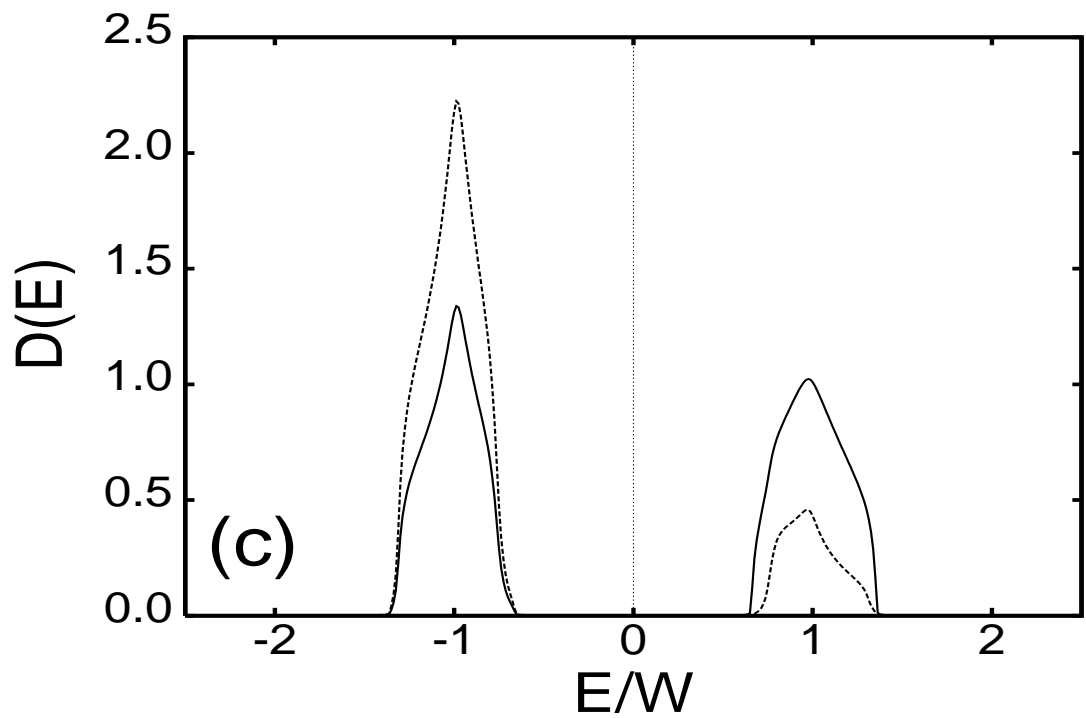

\caption{The densities of states (full line) and the local density of
states on $U$ site
(dashed line) for $U/W=-2.0$ and $n=1$  for different concentrations
$c$: $c=0.4$ ($a$ - metal), $c=0.5$ ($b$ - insulator)
and $c=0.6$ ($c$ - superconductor).
}  
\end{figure}

\begin{figure}
\caption{The order parameter $\bar \chi$ ($a$), the local pairing
potential
on $U$ site $\Delta_U$ ($b$)
and the local charge on $U$ site $n_U$
($c$) and the chemical potential $\mu$ ($d$)
 plotted versus concentration $c$ for $U/W=-2.0$ and several values of
$n$
($n=$ $1.8$, $1.6$, $1.4$, $1.2$, $1.0$, $0.8$, $0.6$, $0.4$, $0.2$ --
the
direction of $n$ changing is
pointed out by the arrow).  
}  
\end{figure}

\begin{figure}
\caption{The order parameter $\bar \chi$ ($a$), the local pairing
potential
on $U$ site $\Delta_U$ ($b$)
and the local charge on $U$ site $n_U$
($c$) and the chemical potential $\mu$ ($d$)
 plotted versus concentration $c$ for $U/W=-0.5$ and several values of
$n$
($n=$ $1.8$, $1.6$, $1.4$, $1.2$, $1.0$, $0.8$, $0.6$, $0.4$, $0.2$ --
the
direction of $n$ changing is
pointed out by the arrow).  
}  
\end{figure}

\begin{figure}
\caption{The quasiparticle densities of states ($a$) and the local
densities
of states on $U$ site ($b$)
for $U/W=-0.5$, $n=0.4$  and  several
values of $c$ specified
in the figures.
}  
\end{figure}

\begin{figure}
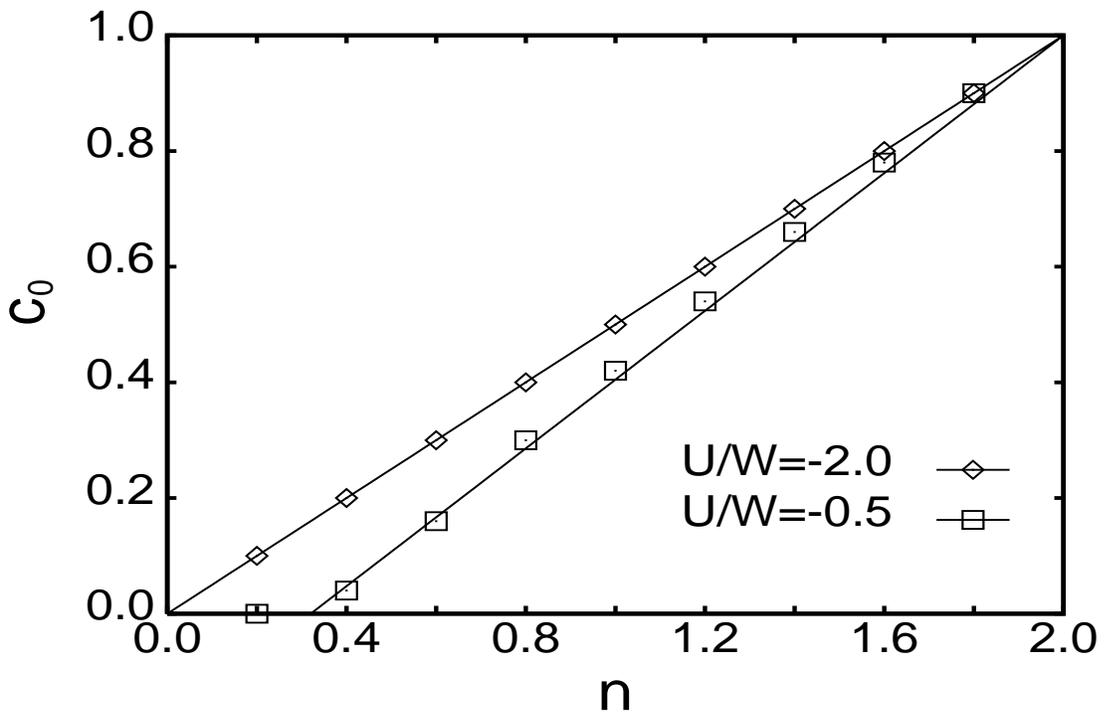

\caption{The critical concentration of negative centers $c$ versus band
filling $n$
for $U/W=-2.0$ and $-0.5$.
}
\end{figure}
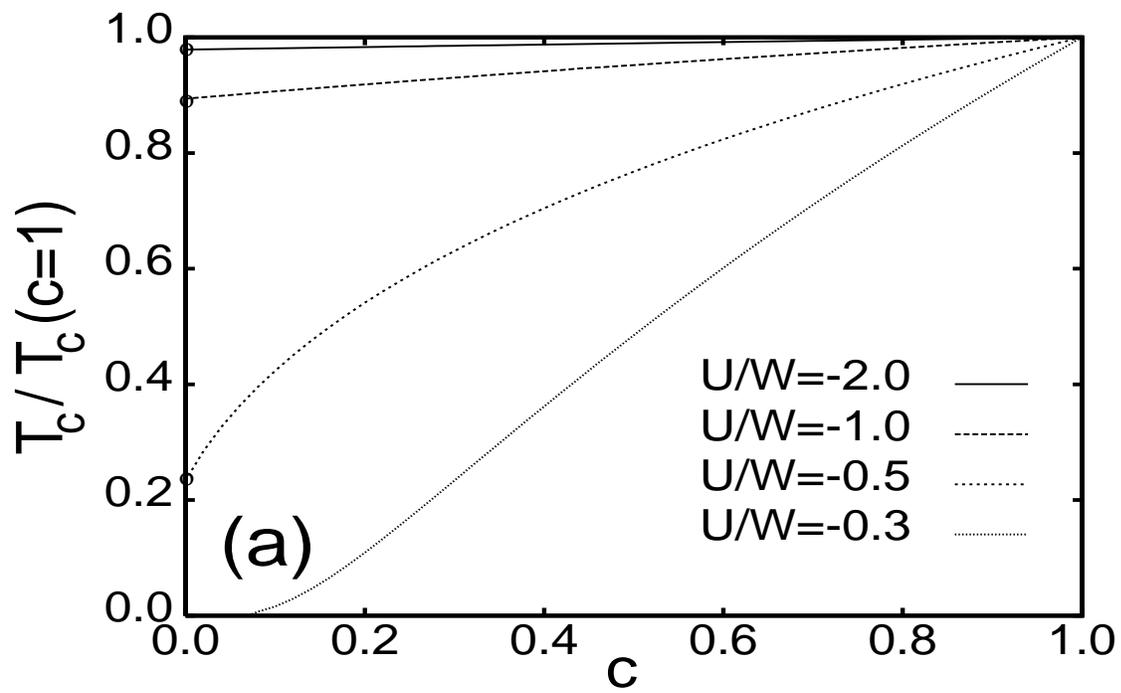
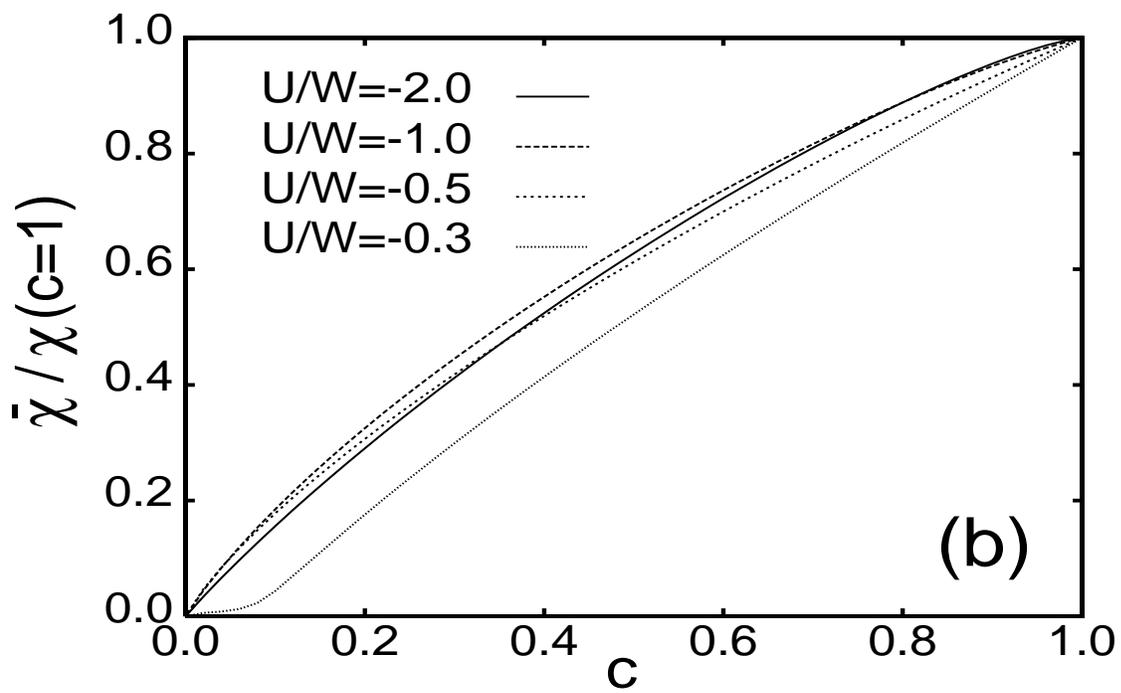
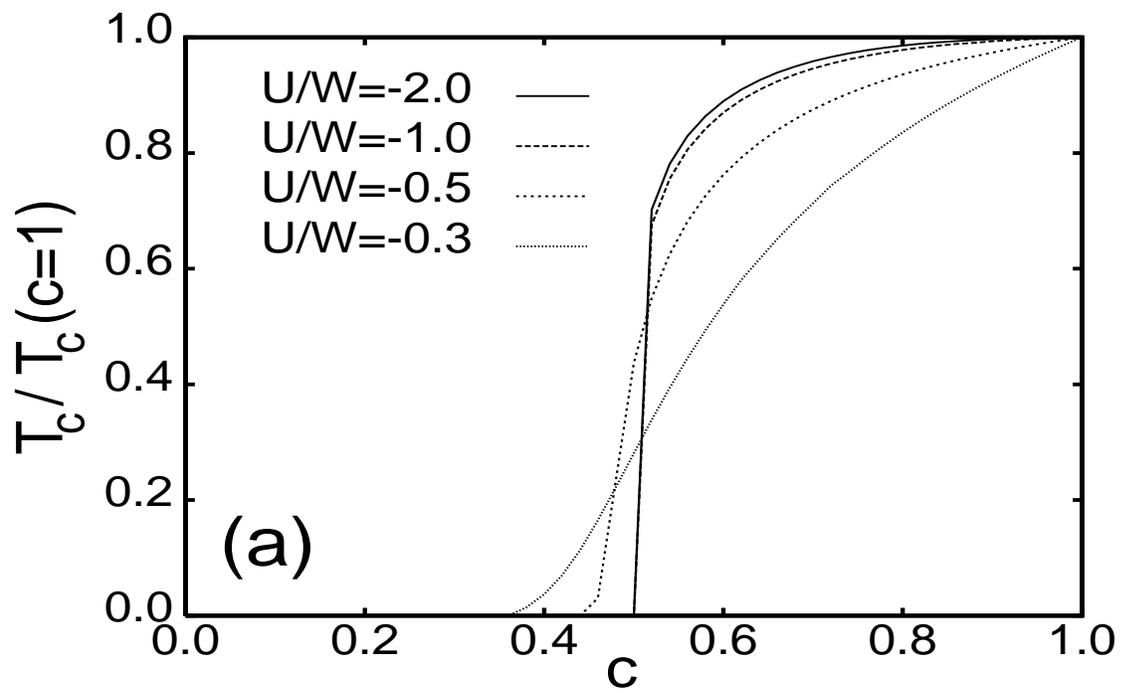
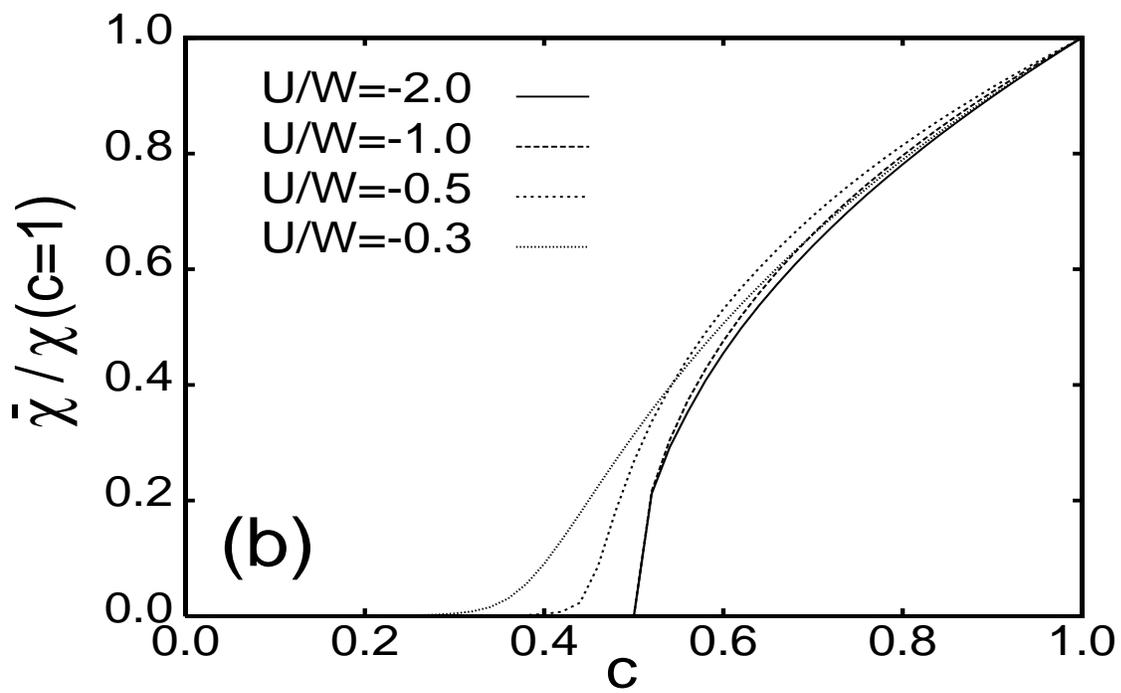
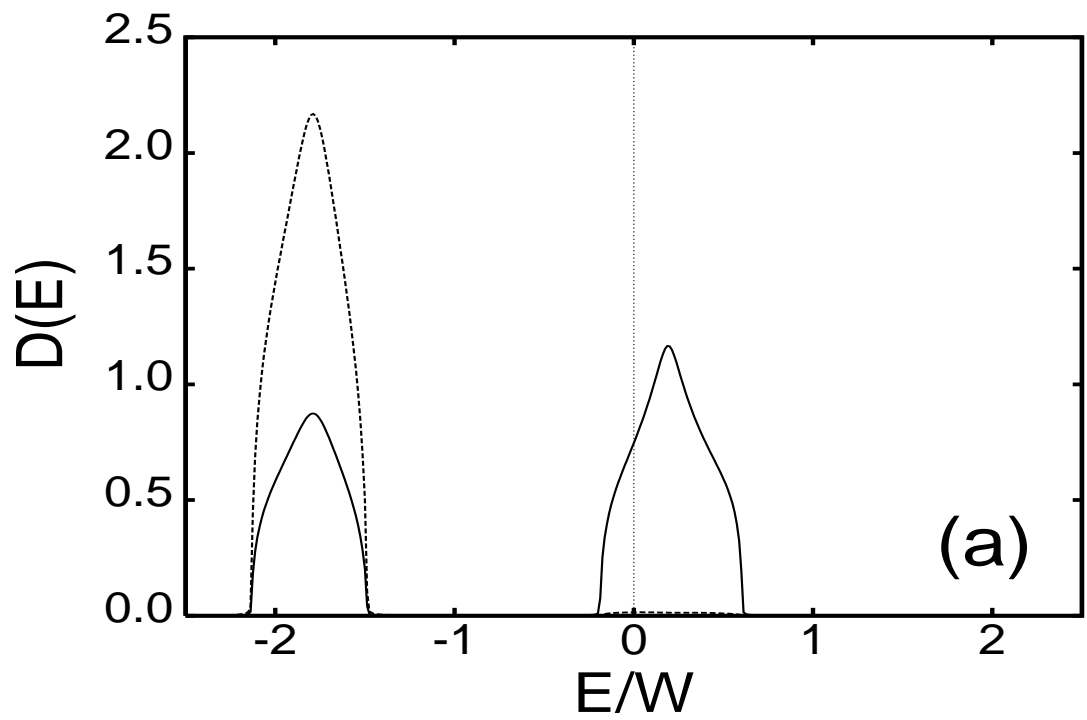
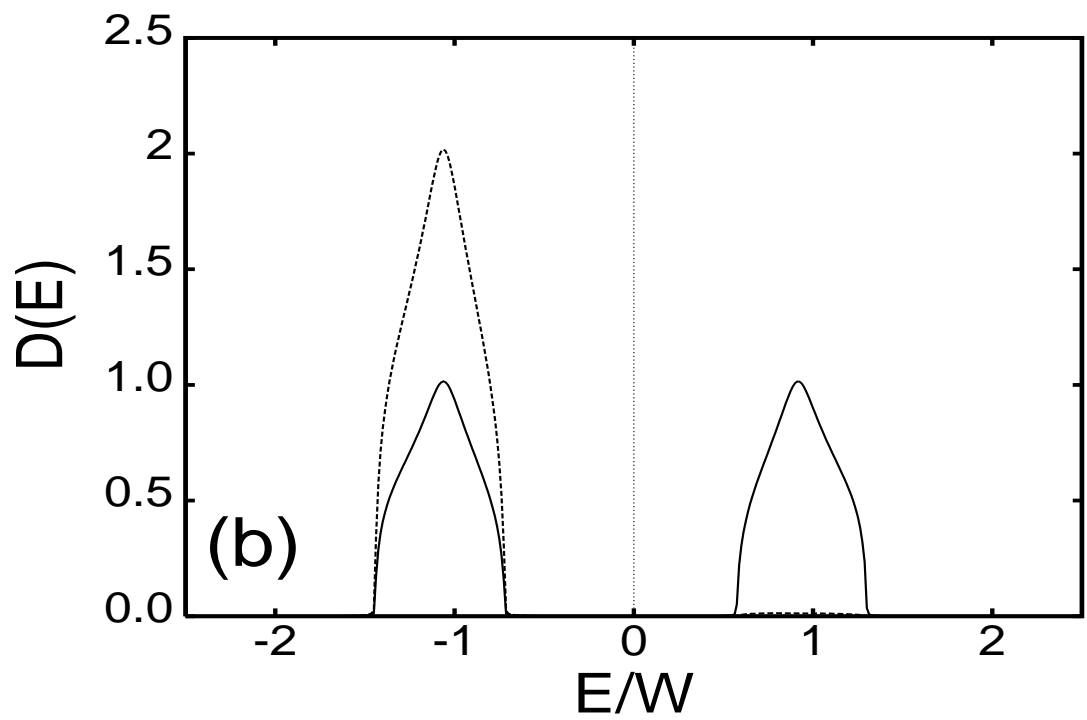
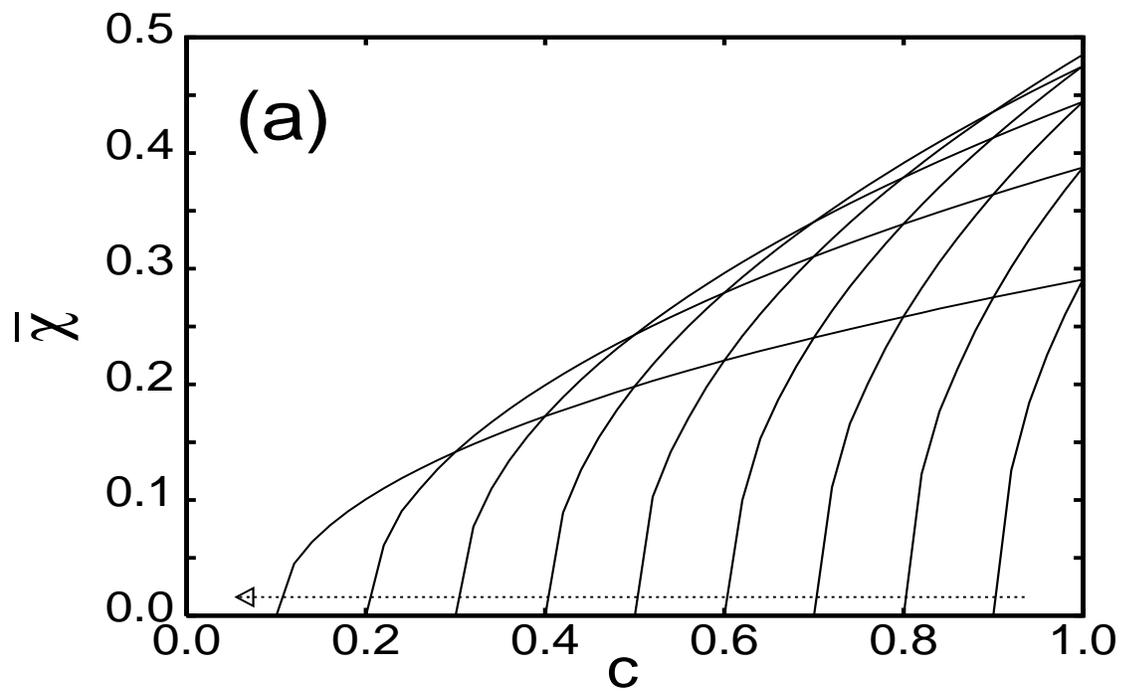
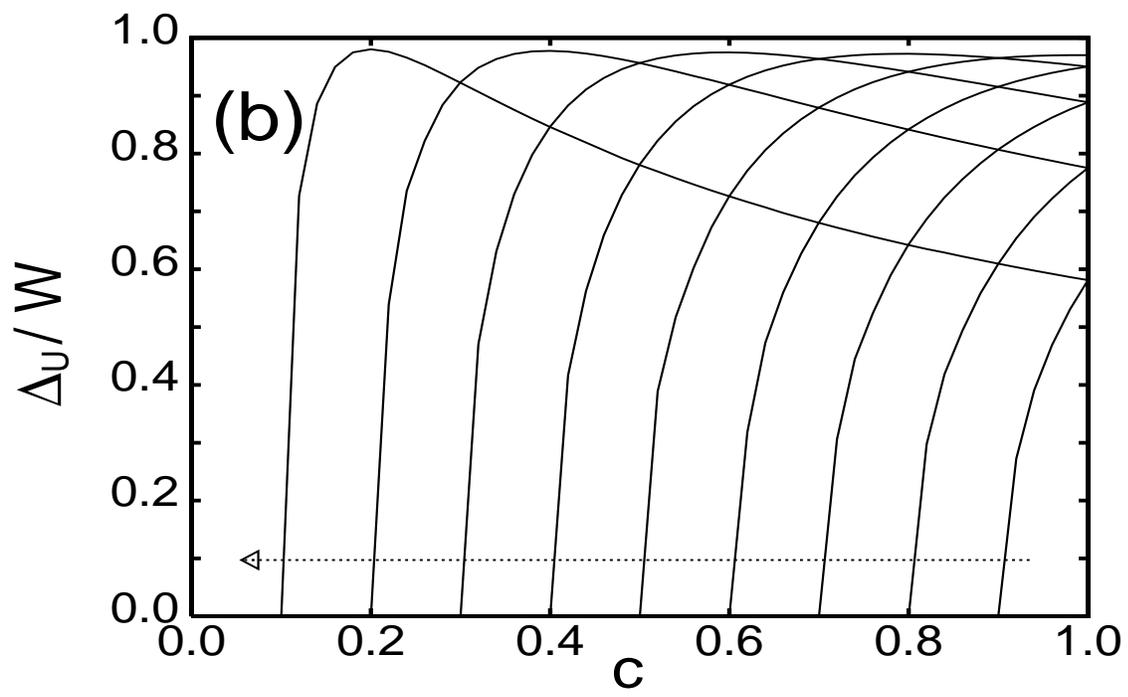
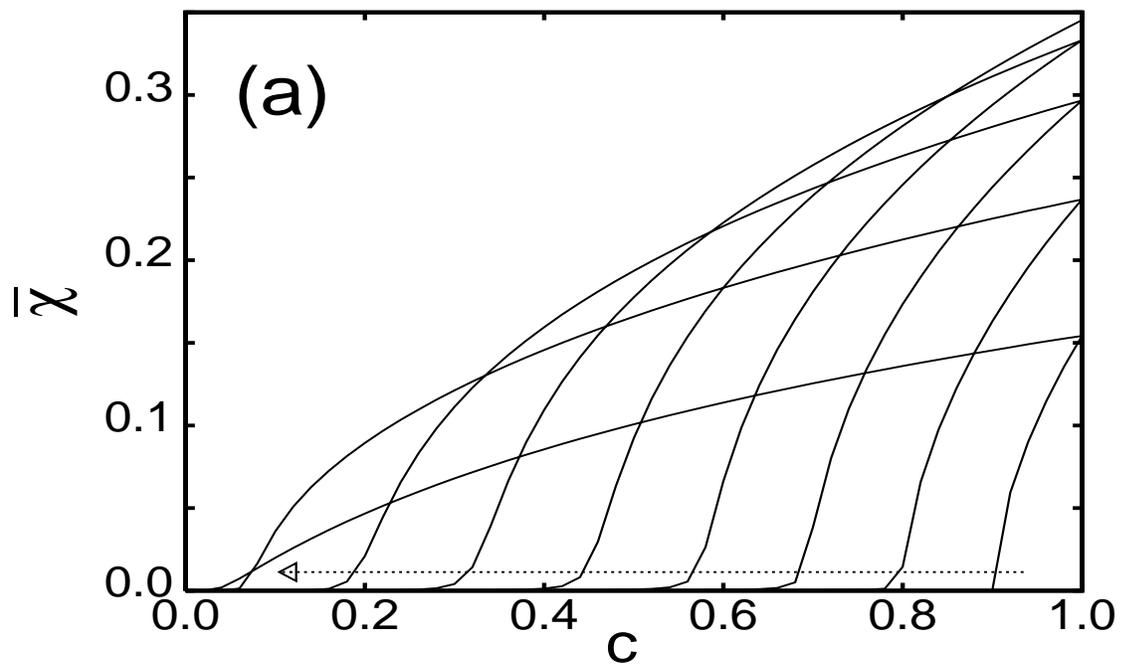
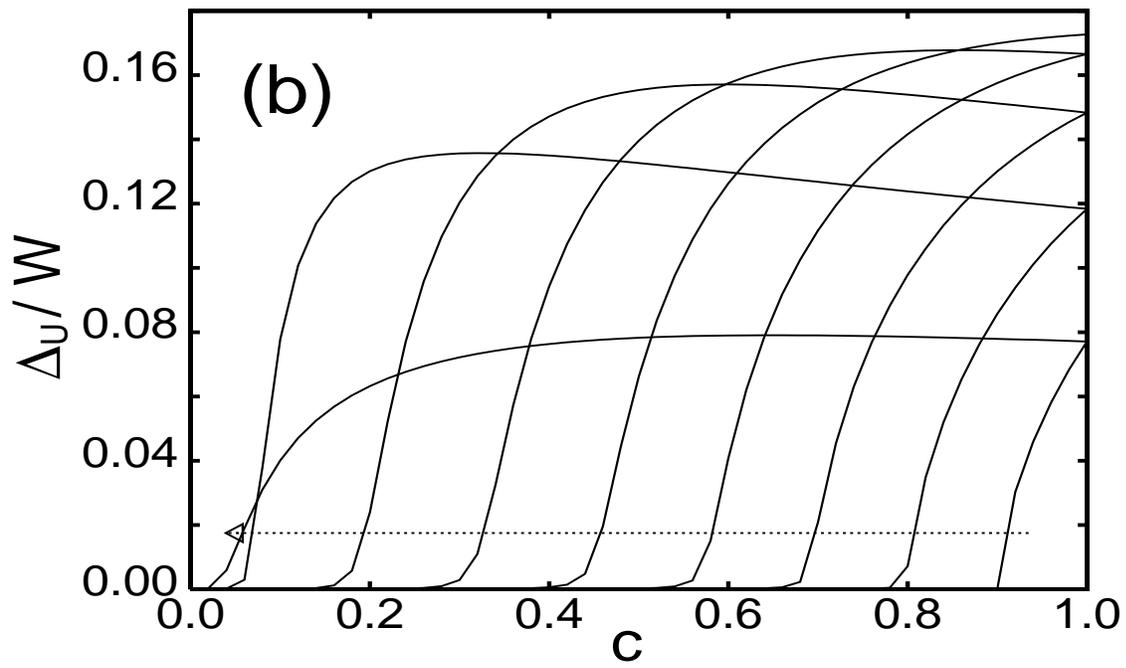
\end{document}